# Evolvable Systems for Big Data Management in Business

R. McClatchey, A. Branson, J. Shamdasani
UWE Bristol, Coldharbour Lane, Frenchay
Bristol BS161QY, UK
{richard.mcclatchey, andrew.branson,
jetendr.shamdasani}@cern.ch

P. Emin
Agilium M1i, Esplanade Augustin Aussedat
74960 Cran-Gevrier, Annecy France
Patrick.emin@agilium.com

*Abstract*— Big Data systems are increasingly having to be longer lasting, enterprise-wide and interoperable with other (legacy or new) systems. Furthermore many organizations operate in an external environment which dictates change at an unforeseeable rate and requires evolution in system requirements. In these cases system development does not have a definitive end point, rather it continues in a mutually constitutive cycle with the organization and its requirements. Also when the period of design is of such duration that the technology may well evolve or when the required technology is not mature at the outset, then the design process becomes considerably more difficult. Not only that but if the system must inter-operate with other systems then the design process becomes considerably more difficult. Ideally in these circumstances the design must also be able to evolve in order to react to changing technologies and requirements and to ensure traceability between the design and the evolving system specification. For interoperability Big Data systems need to be discoverable and to work with information about other systems with which they need to cooperate over time. We have developed software called CRISTAL-ISE that enables dynamic system evolution and interoperability for Big Data systems; it has been commercialised as the Agilium-NG BPM product and is outlined in this paper.

## I. BACKGROUND

In the age of the Cloud and Big Data [1], systems must be increasingly flexible, reconfigurable and adaptable to change in order to respond to enterprise demands. Designing systems to cater for evolution is thus becoming critical to their success. To be able to cope with change, systems must have the capability of design reuse and the ability to adapt as and when necessary to changes in requirements. In addition the systems must capture historical information about how the system has evolved, increasingly termed provenance [2], to enable data traceability. Allowing systems to be self-describing is one way to facilitate this and there have been some advances recently in systems design which enables us to start to build self-describing systems based on the concepts of meta-data, metamodels and ontologies.

Research efforts to tackle problems of system evolution have included design versioning [3], 'active' object models [4] and schema versioning [5]. However, none of these approaches enable the design of existing running systems to be changed on-the-fly and for those changes to be reflected in a new running version of that design nor for external systems to discover how they may inter-operate with the existing system. Businesses expect systems to be agile in nature, to be able to cope with heterogeneity between systems, to have inter-business synchronization and to be responsive to changes in user requirements so that they can evolve over time as the user needs change. The skills to design such complex systems that can facilitate inter-operation and system discovery as a consequence of how they have been designed currently resides in academia. These skills now need to be transferred to commercial developers to enable future production software to be developed.

What industry needs are the skills to develop semantically rich, open, scalable and flexible Big Data systems whose descriptions can be discovered by other cooperating systems so that they can coexist and inter-operate. The UK government has recognised this and has launched an initiative to fund government directed collaborative research into metadata and its applications for use in improved business processes (See: http://www.innovateuk.org/content/competition/metadata-increasing-the-value-of-digital-content-f.ashx).

Researchers in the Centre for Complex Cooperative Systems (CCCS) at the University of the West of England (UWE) have implemented the CRISTAL-ISE system (as described in [6]), based on data and meta-data descriptions that enable systems to dynamically reconfigure and to have system descriptions managed alongside provenance data. CRISTAL-ISE has been used as the basis of the Agilium-NG product (www.agilium.com [7]) to provide the ability to capture system semantics and have those semantics 'discovered' by external systems. Agilium-NG, produced by the M1i company based in Annecy, France, supports Big Data business process management (BPM [8]) and the integration and co-operation of multiple processes especially in business-to-business applications. The product addresses the harmonisation of business processes by the use of the CRISTAL-ISE kernel so that multiple, potentially heterogeneous, processes can be integrated with each other and have their workflows tracked in the database.

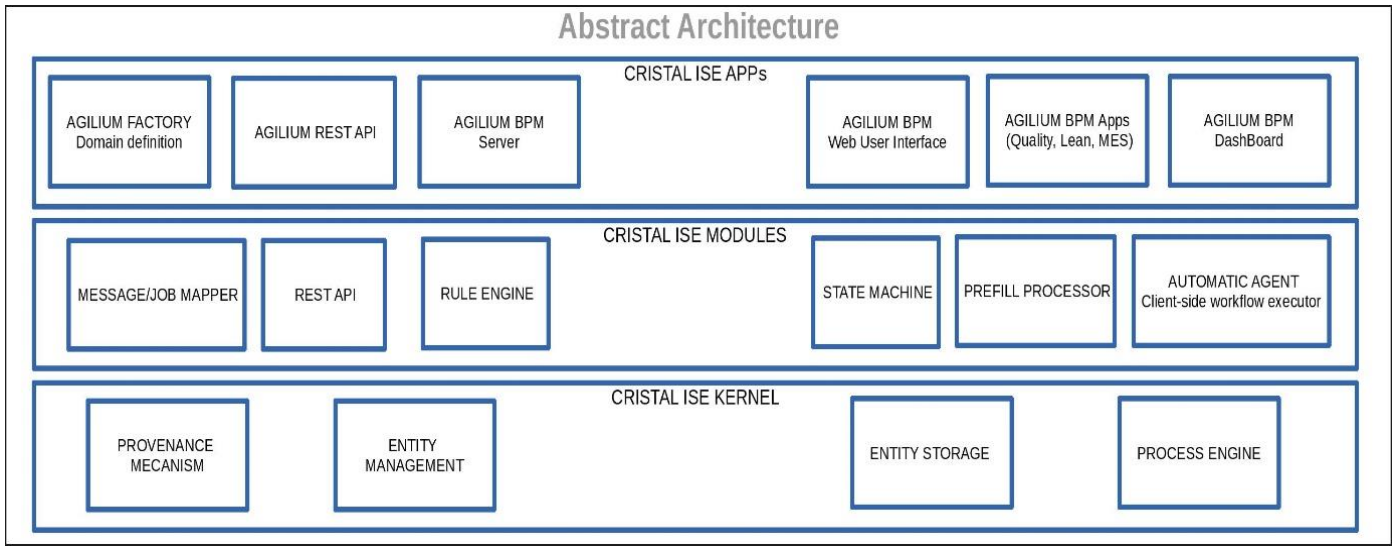

Fig. 1. Agilium-NG's use of the CRISTAL-ISE Kernel.

As a direct result of it being based on CRISTAL-ISE, Agilium-NG is the only existing BPM software on the market that is able to modify an on-going process, in a graphical form and in real time. Without interrupting the business processes, the user can make modifications directly and graphically of any process parameter (flow, roles, GUIs etc). This function guarantees company flexibility by allowing it to react rapidly to external events which have not been modelled without having to review or complicate the model itself.

## II. CRISTAL-ISE AND DESCRIPTION-DRIVEN SYSTEMS

Description-driven systems (DDS) design involves identifying and abstracting, at the outset of the design process, all the crucial elements (such as business objects, processes, lifecycles, goals, agents and outputs) in the system under consideration and creating high-level descriptions of these elements which are then stored in a model, dynamically modified and managed separately from their instances. In many ways adhering to a description-driven approach means following very closely the original principles of pure object-oriented design especially those of reuse, abstraction, and loose coupling. A DDS makes use of so-called meta-objects to store domain-specific system descriptions, which control and manage the lifecycles of meta-object instances, or domain objects [6]. In a DDS, descriptions are managed independently to allow the descriptions to be specified and to evolve asynchronously from particular instantiations of those descriptions. Separating descriptions from their instantiations allows new versions of items to coexist with older versions. This separation is essential in handling the complexity issues facing many Big Data computing applications and allows the realization of interoperability, reusability and system evolution since it gives a clear boundary between the application's basic functionalities from its representations and controls.

The CRISTAL-ISE project was initiated to facilitate the management of the engineering data collected in the construction of CERN'S experiment, CMS [9] at the Large Hadron Collider (LHC). CRISTAL-ISE is a distributed product data and workflow management system which makes use of an OO-like database for its repository, a multi-layered architecture for its component abstraction and dynamic object modelling for the design of the objects and components of the system [10].

The DDS approach has been followed to handle the complexity of such a data-intensive system and to provide the flexibility to adapt to changing usage scenarios. The design of CRISTAL-ISE required adaptability over extended timescales for schema evolution, interoperability and for reusability. In adopting a DDS approach the separation of object instances from object description instances was needed. This abstraction resulted in the delivery of a three layer DDS architecture [6]. Our CRISTAL-ISE approach is similar to the familiar model-driven design concept [11], but differs in that the descriptions and the instances of those descriptions are implemented as objects (Items) and importantly, they are implemented and maintained using exactly the same internal model.

## III. AGILIUM-NG USE OF CRISTAL-ISE

The Agilium-NG system was allowed to be open in design; the elegance of its design was not compromised by being viewed from one or several application-led standpoints. Agilium-NG integrates the management of data coming from different sources and unites BPM with Business Activity Monitoring (BAM) [12] and Enterprise Application Integration [13] through the capture and management of their designs in the CRISTAL-ISE Kernel.

Using the facilities for description and dynamic modification in CRISTAL-ISE, Agilium-NG is able to provide modifiable and reconfigurable Big Data business workflows. It uses the description-driven nature of the CRISTAL-ISE model to act dynamically on process instances already running and can thus intervene in the actual process instances during execution (see Figure 1).

These processes can be dynamically (re-)configured based on the context of their execution without compiling, stopping or starting the process and the user can make modifications directly and graphically of any process parameter. Thus the

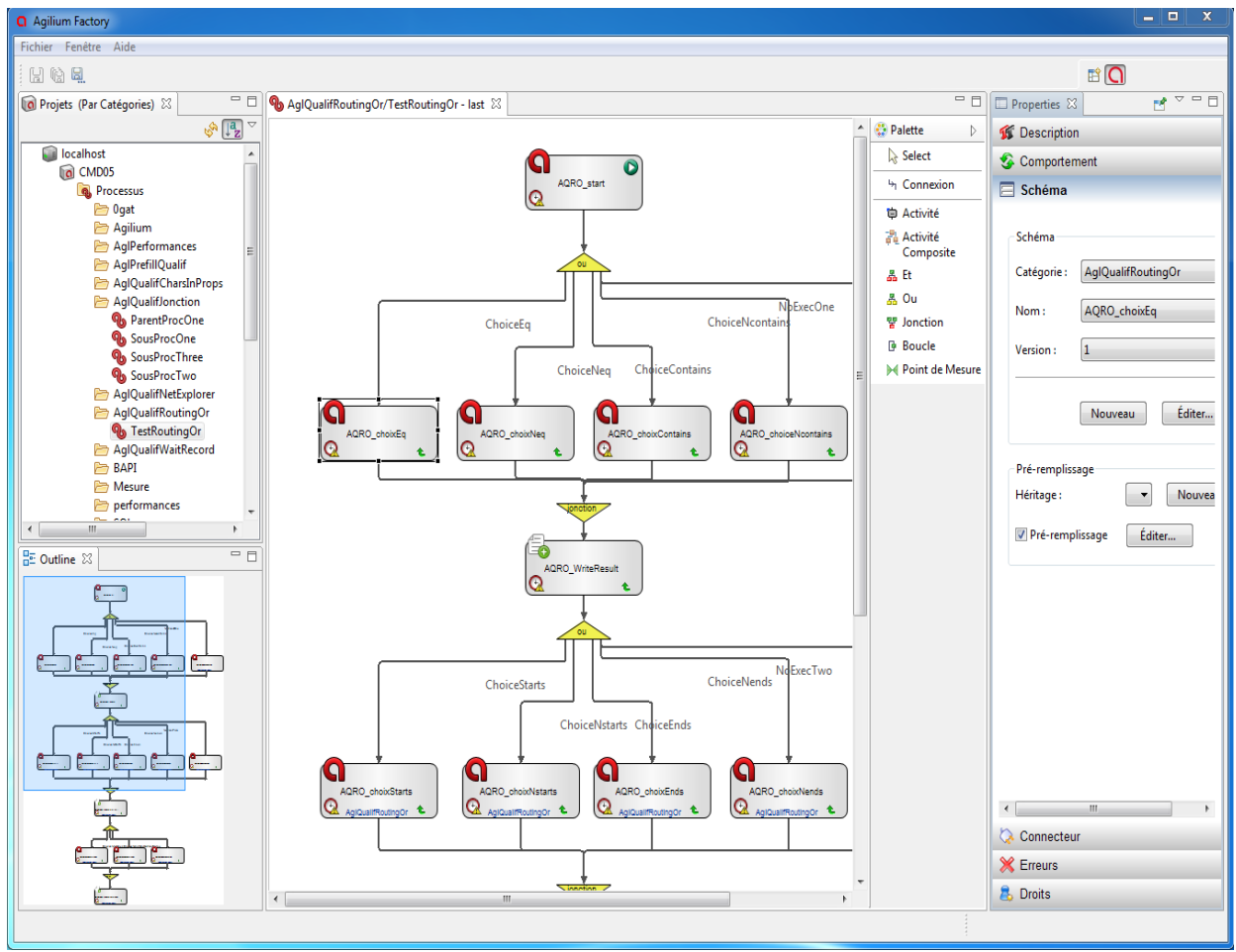

Fig.2. Screenshot of the Agilium-NG Factory Application

Agilium system provides the level of flexibility for organizations to be agile in responding to ongoing changes in Big Data samples and analytics required by cyber-enterprises, with functionality derived from use of CRISTAL-ISE.

The Agilium-NG Server has several domain extensions and support for additional protocols added. The user interface (UI) components are the Agilium Web component, the Agilium Supervisor GUI and the Agilium Factory [5] (see screenshot in Figure 2). The Agilium Web is a web application based on J2EE and running within Tomcat as the container. This is where users can browse the currently active jobs and different instances of business processes. The list of jobs available to a user are constrained by their individual roles (for example, administrator). The web UI also allows users to complete manual activities. The supervisor GUI component of Agilium is derived from the original Java Swing CRISTAL-ISE GUI, and is used by administrators of the system to be able to design and debug workflows and for general system management.

The key component in Agilium is known as the Factory. The Factory is a full Eclipse based application which has a modern UI and allows M1i's users to create and manage their own CRISTAL-ISE based workflows. The workflow definitions are captured in the DDS model of CRISTAL-ISE (the Kernel) and are available for querying and instantiation for the CRISTAL-ISE and Agilium-NG modules (see Figure 2).

The major benefit to Agilium in the use of CRISTAL-ISE is in provenance capture and recording of their Business Process Modelling (BPM) workflow executions. Within the Agilium product, the provenance model is identical to the provenance model of CRISTAL-ISE where Events are generated and stored. As stated previously, all models are created at runtime. This means that all BPM workflows developed within Agilium are stored and versioned (and thus their traceability, or provenance, is recorded). This allows users to return at a later date and view previous versions of the BPM models, fix bugs, or to extend their previous BPM workflows in a new design.

In Agilium-NG the CRISTAL-ISE kernel can capture information about the application area in which a particular instance of Agilium-NG is being used, such as the domains of finance, production control, retail management, etc. This can in principle allow usage patterns to be described and captured, roles and agents to be defined on a per-application basis and rules and outcomes specific to particular user domains to be managed. In turn this enables multiple instances of Agilium-NG to discover the semantics needed for them to inter-operate and to exchange data. This will have a profound effect in terms of business-to-business operation and the ease of configuration and maintainability of systems involving multiple instances of Agilium-NG.

With CRISTAL-ISE we can realise this functionality by including semantics and semantic descriptions in the Agilium-NG product. Using this we can enable business to business (B2B) collaboration by, for example, allowing one company using Agilium-NG to discover information from another company also using Agilium-NG and thereby to facilitate cooperation. As an example one company supplying spark plugs in the automotive industry to another company (Engine assembly) could use the CRISTAL-ISE enhanced version of Agilium-NG to discover information it requires to enable the automatic supply of the parts e.g. by tracking versions of the parts that are required and used in Peugeot car models. This would consequently provide B2B functionality at the production level and ease inter-company collaboration.

## IV. Discussion, conclusions and Future work

The study described in this paper has demonstrated the benefits of a self-describing description-driven design. It has shown that describing a Big Data system explicitly and openly in a model from the outset of a project enables the developers to change aspects of it responsively as users' requirements evolve. This enables seamless transition from version to version with (virtually) uninterrupted system availability; the consequent continuity of function facilitates full traceability of data and processes throughout the system lifecycle. In practice we have found that many system elements have gained in conceptual simplicity and consequent ease of management thanks to loose typing and the adoption of a unified approach to their online manipulation: activities/scripts and their methods, member types and instances, properties and primitives, items and collections, and outcome schemas and views. One logical consequence of providing such a unified design and simplicity of management is that the CRISTAL-ISE software can be used for a wide spectrum of application domains whether from the research domain or in practical application in industry (as shown by the examples described in this paper). Following the principles of object-oriented design the description-driven approach encourages reuse of code, configuration data and scripts/methods. Indeed, the description-driven design approach takes this one step further and provides reuse of meta-data, design patterns and maintenance of items and activities (and their descriptions). Practically this results in a higher level of control over design evolution and simpler implementation of system improvements and easier maintenance cycles

Future work is being carried out to model domain semantics e.g. the specifics of a particular application domain such as healthcare, public sector, finance, and aerospace. This will essentially transform CRISTAL into a self-describing model execution engine, making it possible to build applications directly on top of the design, largely without code generation. The design will be the framework for all of the application logic – without the risks of misalignment and subsequent loss that code generation can bring – and for CRISTAL to be configured as needed to support any Big Data application logic. What this means is that the CRISTAL kernel will be able to capture information about the application area in which a particular instance is being used. This will allow usage patterns to be described and captured, roles and agents to be defined on a per-application basis, and rules and outcomes specific to particular user domains to be managed. This will enable multiple instances of CRISTAL to discover the semantics required to inter-operate and to exchange data.

It is planned to investigate how the semantics of CRISTAL-ISE items and agents could be captured in terms of ontologies and thus mapped onto or merged with existing ontologies for the benefit of new domain models. The emerging technology of Big Data analytics and cloud computing and its application in complex domains, such as medicine and healthcare, provide further interesting challenges. To support this in Q4 2014 a version of CRISTAL-ISE was released to the public as Open Source under the LGPL V3.0 licensing scheme (see www. http://cristal-ise.github.io/).

## Acknowledgment

The authors wish to highlight the support of their home institute across all of the projects that led to this paper. They also acknowledge the support of the European Union for the CRISTAL-ISE project under the 2011-2012 Marie Curie Industry and Academic Pathways Partnership (IAPP) scheme contract number : 285884.